\documentclass[conference]{IEEEtran}
\usepackage{cite}

\ifCLASSINFOpdf
\else
\fi

\usepackage{graphics}
\usepackage{graphicx}
\usepackage{subfigure}

\begin{document}
%
\title{Comprehensive Measurement of Neutron Yield Produced by 62 MeV Protons on Beryllium Target}



%
\author{\IEEEauthorblockN{M.~Osipenko\IEEEauthorrefmark{1},
M.~Ripani\IEEEauthorrefmark{1}, R.~Alba\IEEEauthorrefmark{2}, G.~Ricco\IEEEauthorrefmark{1},
M.~Schillaci\IEEEauthorrefmark{2}, M.~Barbagallo\IEEEauthorrefmark{3}, P.~Boccaccio\IEEEauthorrefmark{4},
A.~Celentano\IEEEauthorrefmark{5},\\
N.~Colonna\IEEEauthorrefmark{3}, L.~Cosentino\IEEEauthorrefmark{2}, A.~Del~Zoppo\IEEEauthorrefmark{2},
A.~Di~Pietro\IEEEauthorrefmark{2}, J.~Esposito\IEEEauthorrefmark{4}, P.~Figuera\IEEEauthorrefmark{2},
P.~Finocchiaro\IEEEauthorrefmark{2},\\
A.~Kostyukov\IEEEauthorrefmark{6}, C.~Maiolino\IEEEauthorrefmark{2},
D.~Santonocito\IEEEauthorrefmark{2}, V.~Scuderi\IEEEauthorrefmark{2}, C.M.~Viberti\IEEEauthorrefmark{5}}
\IEEEauthorblockA{\IEEEauthorrefmark{1}INFN, sezione di Genova, 16146 Genova, Italy}
\IEEEauthorblockA{\IEEEauthorrefmark{2}INFN, Laboratori Nazionali del Sud, 95123 Catania, Italy}
\IEEEauthorblockA{\IEEEauthorrefmark{3}INFN, sezione di Bari, 70126  Bari, Italy}
\IEEEauthorblockA{\IEEEauthorrefmark{4}INFN, Laboratori Nazionali di Legnaro, 35020 Legnaro, Italy}
\IEEEauthorblockA{\IEEEauthorrefmark{5}Dipartimento di Fisica dell'Universit\`a di Genova, 16146 Genova, Italy}
\IEEEauthorblockA{\IEEEauthorrefmark{6}Moscow State University, Moscow 119992, Russia}
}


\maketitle

\begin{abstract}
A low-power prototype of neutron amplifier, based on a 70 MeV, high current proton cyclotron being installed at LNL
for the SPES RIB facility, was recently proposed within INFN-E project. This prototype uses
a thick Beryllium converter to produce a fast neutron spectrum feeding a sub-critical reactor core.
To complete the design of such facility the new measurement of neutron yield from a thick Beryllium target
was performed at LNS. This measurement used liquid scintillator detectors to identify produced neutrons
by Pulse Shape Discrimination and Time of Flight technique to measure neutron energy in the range 0.5-62 MeV.
To extend the covered neutron energy range $^3$He detector was used to measure neutrons below 0.5 MeV.
The obtained yields were normalized to the charge deposited by the proton
beam on the metallic Beryllium target. These techniques allowed to achieve a wide angular coverage
from 0 to 150 degrees and to explore almost complete neutron energy interval.
\end{abstract}


%
\IEEEpeerreviewmaketitle

\section{Introduction}
Despite the advent of alternative sources nuclear fission remains one of the main techniques of energy production. Limited availability of fissile materials and difficulties in nuclear waste disposal
call for research on fast neutron reactors.
The interest for fast neutrons stems from the fact that they induce fission
in broader number of elements while suppressing capture processes leading to high toxicity waste build-up.
In particular, minor actinides like $^{241}$Am, $^{244}$Cm etc., have a fission
threshold of about 0.5 MeV, therefore only fast neutrons can burn them out.
Accelerator Driven System (ADS)~\cite{ADS} represents a promising solution to this problem.
The ADS consists of a fast sub-critical reactor fed by an external neutron source.
External neutron source of sufficiently high intensity can be generated by a particle accelerator or by a fusion reactor.

Inspired by the upcoming installation of a high power proton cyclotron at the Laboratori Nazionali di Legnaro
of INFN for the SPES project~\cite{SPES}, the conceptual design of a low-power ADS prototype
for research purposes has been developed in the framework
of INFN-E project~\cite{infn_e_ads}. It is based on 0.5 mA, 70 MeV proton beam
impinging on a thick Beryllium target and fast sub-critical core, consisting of solid lead
matrix and 60 $UO_2$ fuel elements, enriched to 20\% with $^{235}$U. Both, the production
target and reactor core, are cooled by a continuous helium gas flow. The ADS is expected
to have an effective neutron multiplication factor $k_{eff}=0.946$,
neutron flux $\phi=3\div 6\times 10^{12}$ n/cm$^2$/s and thermal power of 130 kW at 200$^\circ$ C.
Such facility would allow to study the kinetics and dynamics of the fast reactor core,
the burn-out and transmutation of radioactive wastes,
as well as issues related to system safety and licensing.

The choice of the production target material is determined by the low beam energy of
the cyclotron. $^9$Be bombarded by protons, in fact, provides an abundant neutron
source. The design of
the proposed ADS requires an accurate knowledge of neutron yield produced by the 70 MeV
proton beam on the $^9$Be target. The existing data in the given energy range are rather
scarce and incomplete. The integrated yield at 70 MeV was measured in Ref.~\cite{Tilquin05},
while differential yields for few angles were measured at various beam energies
in Refs.~\cite{Waterman79,Johnsen76,Almos77,Heintz77,Meier88,Madey77,Harrison80}.
This lack of data demanded a dedicated measurement of neutron yield produced by the
proton beam on a thick $^9$Be target. The measurement was performed
using the superconducting cyclotron~\cite{lnsrep10} at the Laboratori Nazionali del Sud (LNS) of INFN.

In this article we describe a number of selected topics on detectors used in the present experiment
and on the data analysis. More complete information about this measurement,
the comparison to various simulation codes
and tabulated experimental data can be found in Ref.~\cite{nima2013}.

\section{Detector Description}
Neutron detectors for the present experiment were made as 4 cm long Aluminum cylindrical cells filled with liquid scintillator.
In order to measure neutron yields at two different distances from the target,
keeping the same solid angle, two different cell diameters were chosen.
Four cells, used as far detectors, had the diameter of 4.6 cm (large detectors)
and other four, used as near detectors placed at half distance from the target,
had a diameter of 2.3 cm (small detectors).

\begin{figure}[!t]
\begin{center}
\includegraphics[bb=2cm 6.5cm 20cm 23cm, scale=0.35]{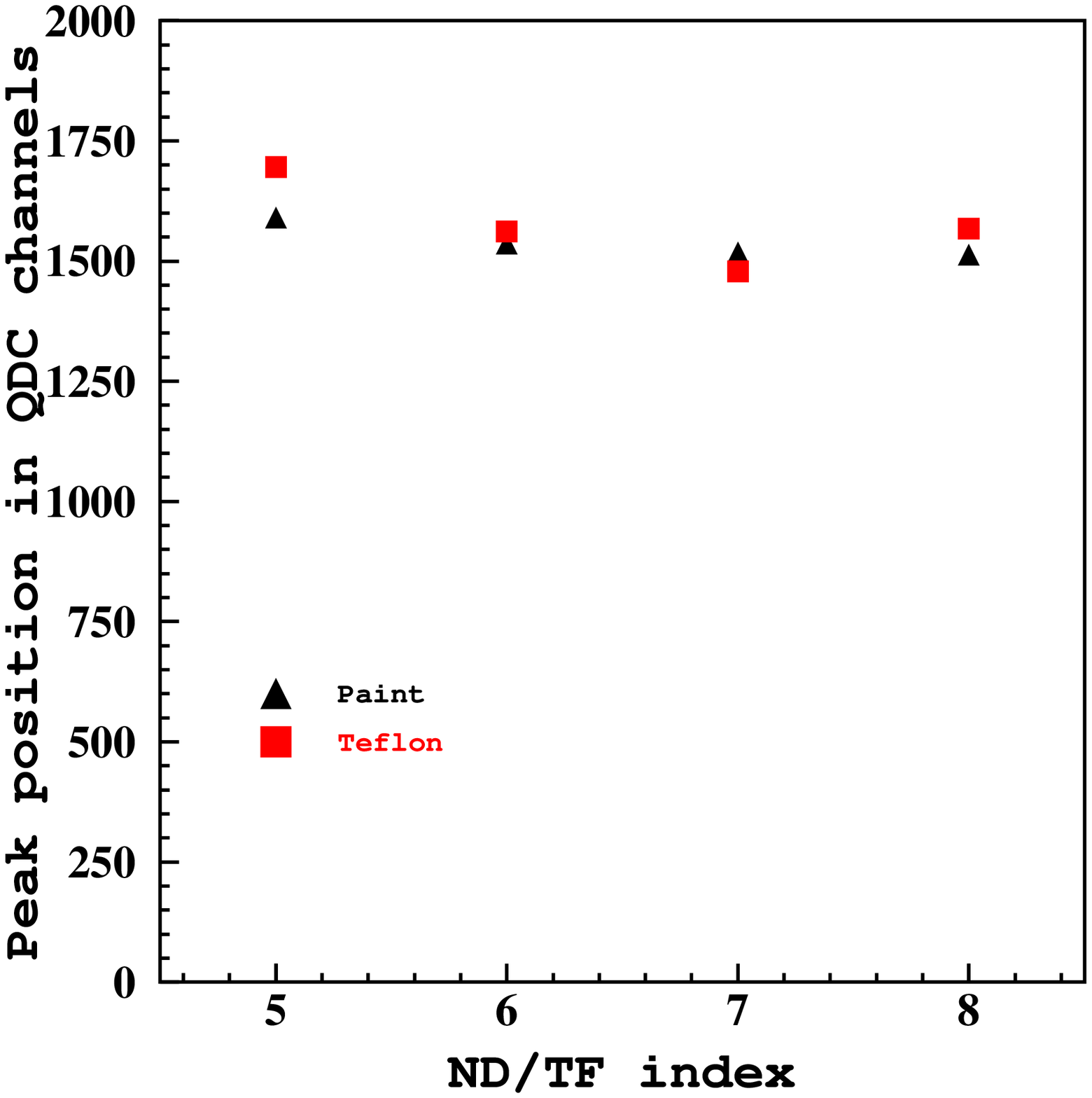}
\includegraphics[bb=2cm 6.5cm 20cm 24cm, scale=0.35]{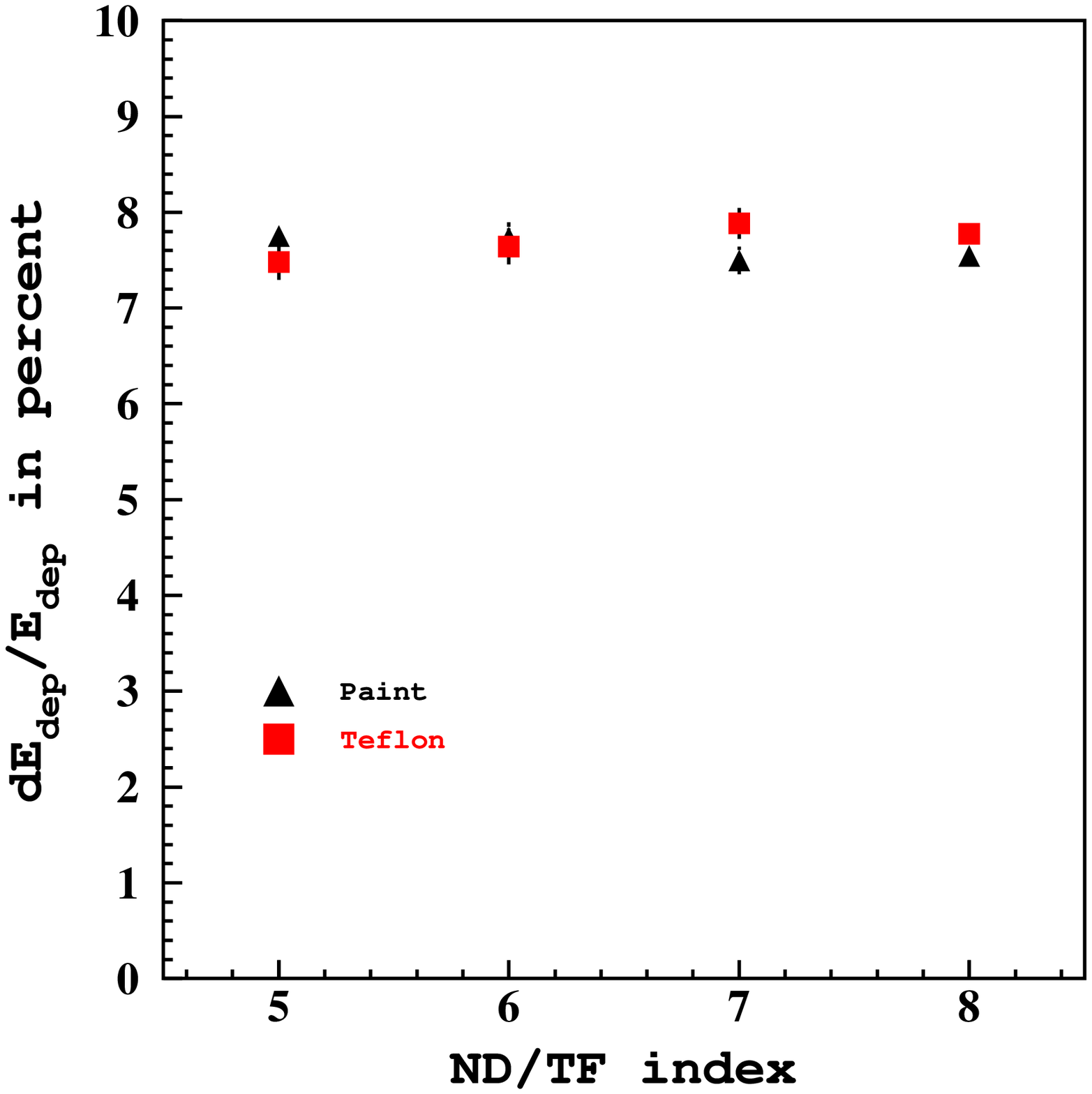}
\caption{\label{fig:cmp_teflon} The position (upper)
and the relative width (lower) of $^{137}Cs$ backward Compton
scattering peak:
black triangles - cells with EJ520 paint,
red squares - cells with Teflon reflector.}
\end{center}
\end{figure}

To enhance light yield on photon sensors, the cell Aluminum walls were covered with a light diffuser.
In order to select the best diffuser material a comparison of light yields
of two detector modifications was performed. This was done only for the small
detectors because they were designed for low light yield signals.
To this end small detectors were manufactured in two versions:
with EJ520 paint diffuser and with Teflon reflector.
Backward Compton scattering spectra were measured in the same conditions (same PMT and QDC channel)
using a $^{137}Cs$ source and the position and the width of the main peak were compared.
The results shown in Fig.~\ref{fig:cmp_teflon} demonstrated that
within the systematic precision (of about 0.5\% on the peak width) the two
reflectors were equivalent. This was compatible with a difference
in the number of detected photoelectrons lower than 12\%,
which was close to the expected 17\% from Ref.~\cite{yamawaki:teflon}.

The cells were filled in Ar/N$_2$ atmosphere with EJ301 liquid scintillator,
having a good Pulse Shape Discrimination (PSD, see next section) capability.
The liquid scintillator did not fill the entire cell volume, leaving
a small bubble of Ar/N$_2$ for a possible thermal expansion.
The total amount of scintillator in each cell was measured
by a comparison of full to empty cell weights, in order
to keep a minimum empty expansion volume of 4\%.

Each cell was sealed by a 4 mm thick borosilicate glass, coupled
to light sensor by optical grease.
As light sensors we selected Electron Tubes Enterprise PMTs 9954 with
high linearity voltage dividers C649. The choice was mainly determined
by a good overall quantum efficiency, extending to the red region. This was
particularly important for PSD~\cite{amaldi:psd_spectrum}.

\subsection{Pulse Shape Discrimination Optimizations}
\begin{figure}[!t]
\begin{center}
\includegraphics[bb=2cm 6.5cm 20cm 23cm, scale=0.35]{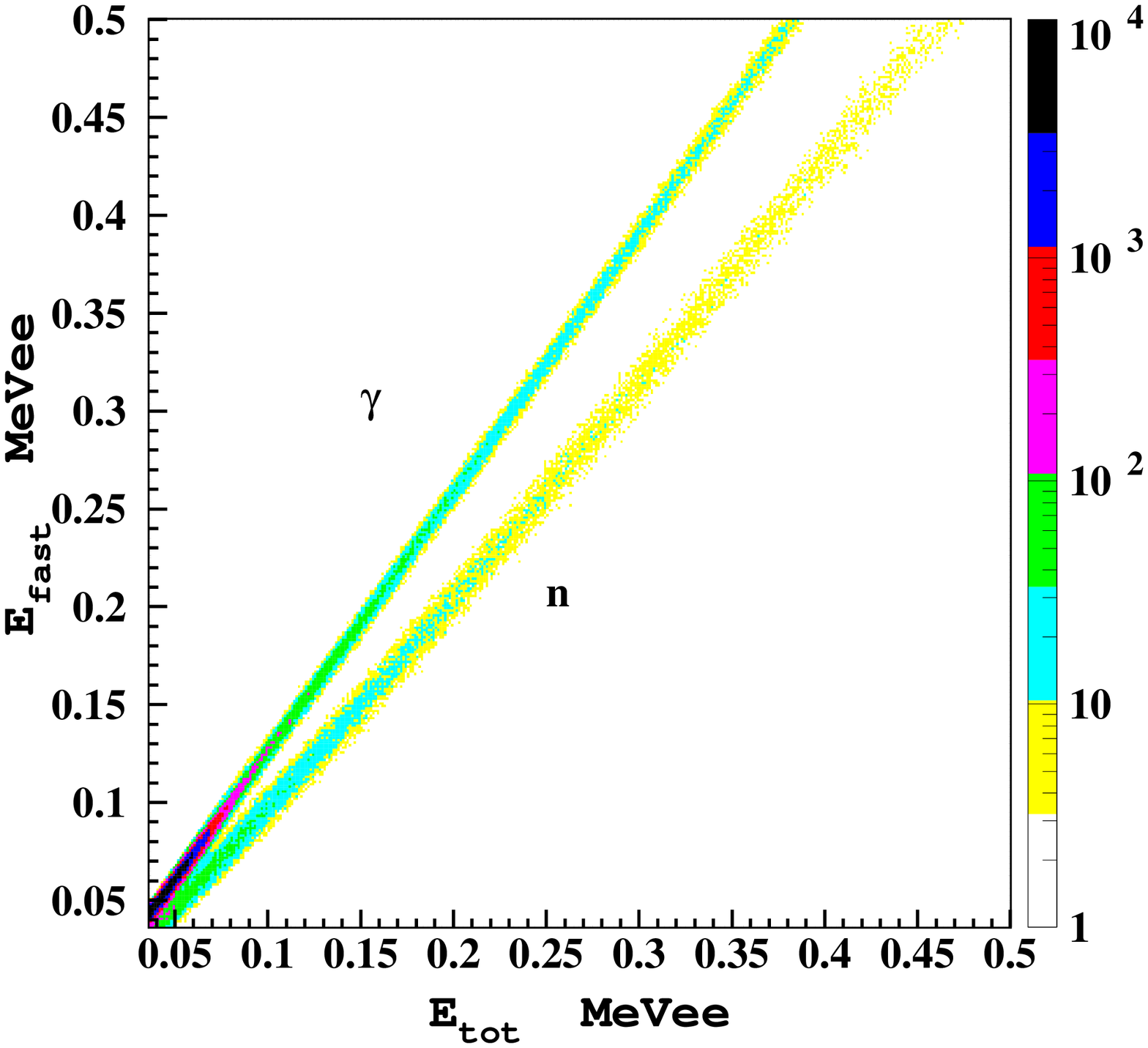}
\includegraphics[bb=2cm 6.5cm 20cm 24cm, scale=0.35]{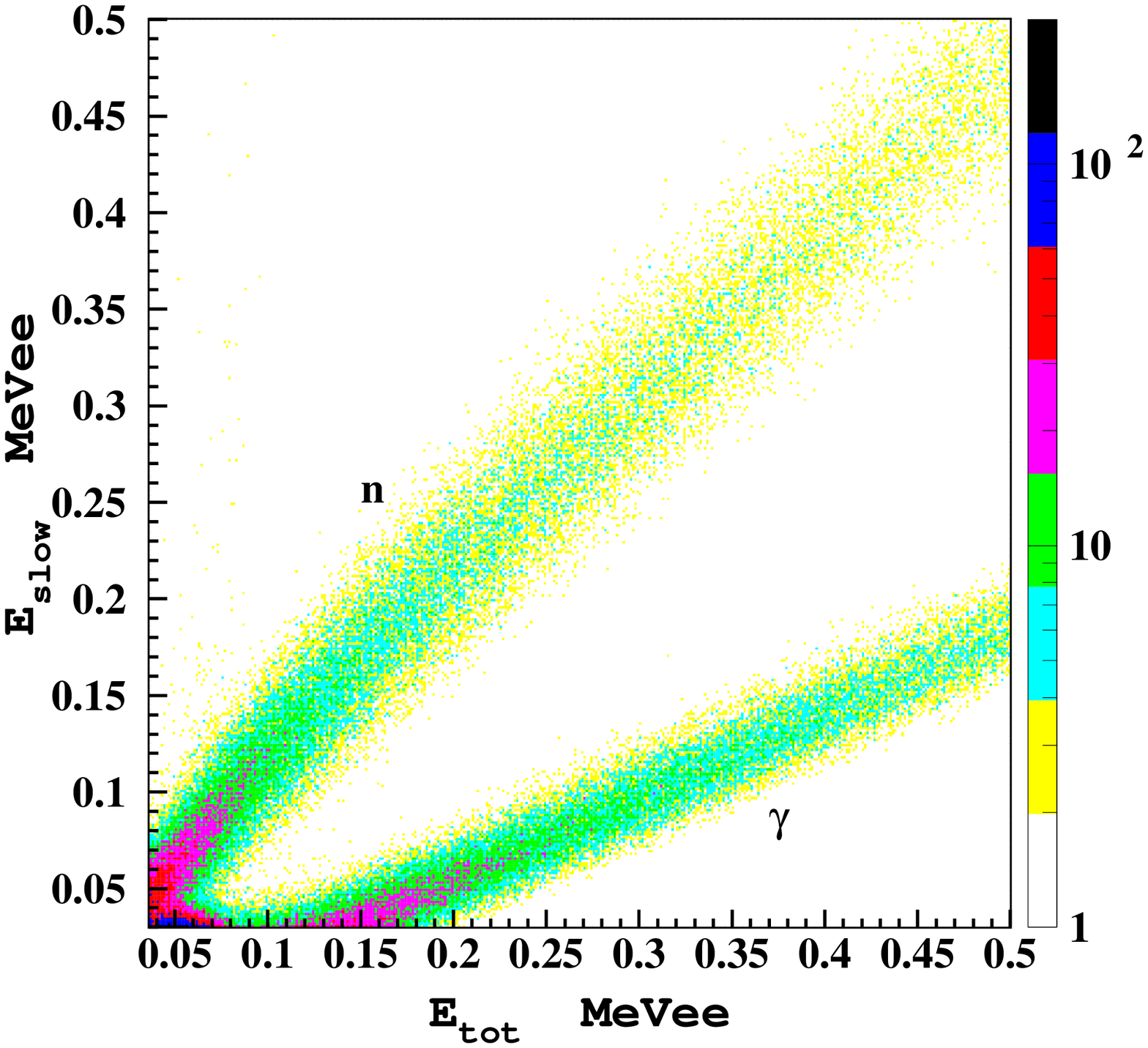}
\caption{\label{fig:psd_edep_part_tot} PSD separation between $\gamma$s and neutrons based on
Fast (upper) and Slow (lower) partially integrated signals.}
\end{center}
\end{figure}

\begin{figure}[!t]
\begin{center}
\includegraphics[bb=2cm 6cm 20cm 23cm, scale=0.35]{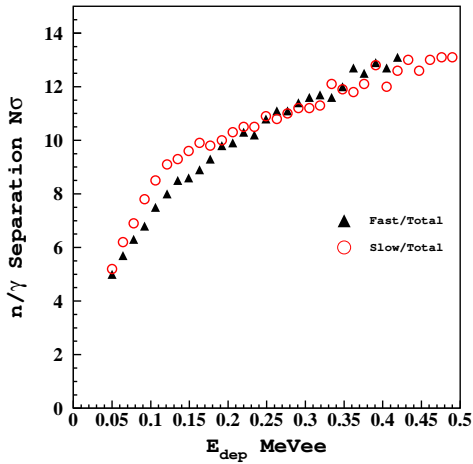}
\caption{\label{fig:psd_fast_slow} PSD separation between $\gamma$s and neutrons for Fast (black)
and Slow (red) partially integrated signals.}
\end{center}
\end{figure}

The proton beam interacting with the Beryllium target produces not only neutrons
but also $\gamma$s. $\gamma$s can be produced at the time of the proton interaction (prompt $\gamma$s),
or when a neutron interacts with surrounding materials (delayed $\gamma$s).
To eliminate the $\gamma$ background Time of Flight (mostly for prompt $\gamma$s)
and PSD were used.
PSD is based on the dependence of the scintillation component superposition
from the local energy deposition density.
In our setup PSD was performed by a comparison of fully integrated signal (Total)
and partially integrated signal. One can choose to integrate either the peak
of the signal (Fast) or its tail (Slow).
For small cells we compared these two methods to identify the best one.
The two obtained PSD plots are shown in Fig.~\ref{fig:psd_edep_part_tot}.
We quantified these separations in terms of number of $\sigma$ as a function
of the total deposited energy as shown in Fig.~\ref{fig:psd_fast_slow}.
We found that the integration of the signal tail was providing the best PSD
in the low deposited energy range, where separation is typically more difficult
due to the lower amount of scintillation light.


\subsection{Threshold Optimization}
To maximize detector efficiency and to reduce the corresponding systematic uncertainty,
a threshold as low as 10 KeVee was desirable, in particular for small cells.
For this purpose the large and small detector signals were handled in different ways.

Large detectors were designed to cover the entire energy range of the experiment up to 60 MeVee.
The signal from PMT of a large detector was split in three parts by a passive splitter:
\begin{itemize}
\item 1/9 were amplified $\times 10$ by a Philips Scientific 771 amplifier and sent to GANELEC FCC8 CFD input,
\item 2/9 were directly connected to a QDC input for the Total deposited energy integration,
\item 2/3 had the cable delay reduced on 50 ns with respect to 2/9,
and were connected to another QDC input for integration
of the Slow part of the signal.
\end{itemize}
In this configuration the QDC Full Scale (FS) was corresponding to the upper limit of PMT voltage divider linearity range ($<$8 V). The Slow QDC charge had a similar magnitude to the Total one, allowing for a good PSD precision.
Adjusting PMT gain to extend the covered deposited energy interval up to 60 MeVee
the minimal discriminator threshold of 8 mV was found to correspond to 200 keVee.

For the small cells, which were designed to cover a deposited energy range
up to 0.5 MeVee, the configuration was different. The PMT signal was first
amplified $\times 6$ and integrated by Ortec 454 TFA with timing constant $\tau_{RC}=20$ ns.
Then it was split in three parts by a passive splitter:
\begin{itemize}
\item 1/9 were connected to QDC input for the Total deposited energy integration,
\item 2/9 were sent to GANELEC FCC8 CFD input,
\item 2/3 were anticipated by 50 ns and were connected to another QDC input for integration
the Slow part of the signal.
\end{itemize}
The main limitation in this case was imposed by the TFA output linearity range ($<$6 V).
Therefore, PMT gains were adjusted by means of $^{137}$Cs source to provide about 1 V signals for 0.5 MeVee
deposited energy. This configuration allowed to achieve thresholds as low as 13 keVee.
Moreover, the integration of the signal by TFA reduced the triggering
on the PMT intrinsic noise to acceptable level ($<20$ Hz).

\section{Data Acquisition System}
A CAMAC based Data AcQuisition system (DAQ) had been developed for this measurement.
The system used LeCroy 8901A GPIB-to-CAMAC interface crate controller for
communication with Linux PC, equipped with National Instruments PCI-GPIB+ board.
GPL Linux-GPIB driver was used as GPIB API.
The overall system speed was limited by the 8901A controller data transfer rate
in block mode of 0.45 Mb/s ($>$1.5 Mb/s for PCI-GPIB+ board).

\begin{figure*}[!ht]
\begin{center}
\includegraphics[scale=0.25, angle=0]{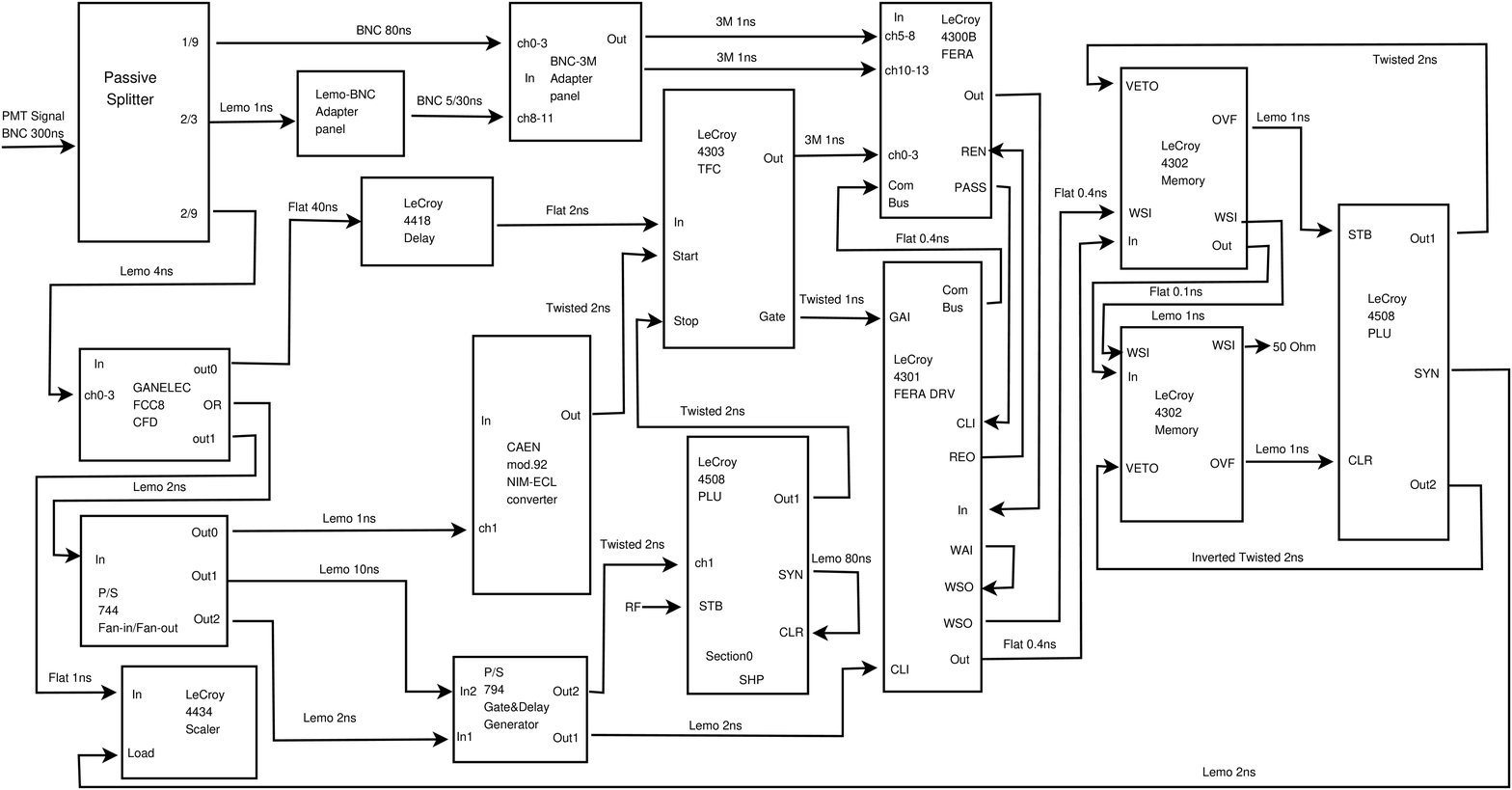}
\caption{\label{fig:daq_scheme} Signal propagation scheme of our CAMAC based DAQ system.}
\end{center}
\end{figure*}

On the CAMAC side we used 11-bit LeCroy 4300B FERA based system with ECL bus readout
on two alternating buffer memories. To improve readout speed we setup ECL bus configuration
connecting 4300B QDC to two 4302 LeCroy memory modules through the LeCroy 4301 FERA Driver.
ECL bus allows for data transfer rate up to 20 Mb/s, well matched to 10 $\mu$s QDC dead time.
An approximate scheme of DAQ system used in this experiment is shown in Fig.~\ref{fig:daq_scheme}.

A hit in one of four detectors connected simultaneously to the DAQ system triggered FERA QDC
integration within 250 ns Gate. The trailing edge of the Gate signal started
the data conversion (8.5 $\mu$s), followed by zero suppression (2.5 $\mu$s)
and data readout through ECL bus (1.5 $\mu$s).
To avoid DAQ blocking on incomplete conversion events
a Clear command was issued after every conversion by means
of the Gate signal, delayed by 13 $\mu$s by a Philips 794 Gate and Delay Generator.

The mechanism of memory switching and readout was based on their Overflow signals
connected to LeCroy 4508 PLU. When the number of written words in memory
reached 15360 it rose Overflow level. The PLU instantly ($<$20 ns) set a veto
on the memory in overflow state and opened the second memory for write access
and so on, alternating data writing from one memory to another.
This is achieved by connecting one memory Overflow signal to PLU Strobe
and the other to its Clear and configuring the two PLU outputs to generate
logical level vetoing the full memory.

The complete readout of 15360 16-bit data words from the memory in block mode
requires about 70 ms. To avoid additional DAQ dead times this
has to be smaller than the memory filling time.
Therefore, the overall event rate of our four-detector system is limited
by $R<16.6$ kHz at 25\% dead time.

\section{Experimental Setup}\label{sec:baf}
A proton beam of 62 MeV from LNS superconducting cyclotron was delivered in the MEDEA experimental hall
with a beam current of 30-50 pA. In these conditions the single detector rate was below a few kHz.
The cyclotron beam structure (RF=40 MHz) was modified by suppressing
four bunches out of five. In this way 1.5 ns wide beam bunches arrived on the target
with period of 125 ns.
The target consisted of a solid, 3 cm thick $^9$Be cylinder with 3.5 cm diameter
installed at the end of a 1 m long last section of beamline,
made of 3 mm thick carbon fiber to minimize secondary $\gamma$s. The target thickness
was chosen to ensure the complete absorption of 70 MeV proton beam.
The electric charge deposited by the beam on the target was measured by a digital current integrator
and used for the absolute normalization of the data. Two different current integrators were used during the experiment for comparison: Ortec 439 and 1000C of Brookhaven Instruments Corporation.

\begin{figure}[!t]
\begin{center}
\includegraphics[scale=0.28, angle=0]{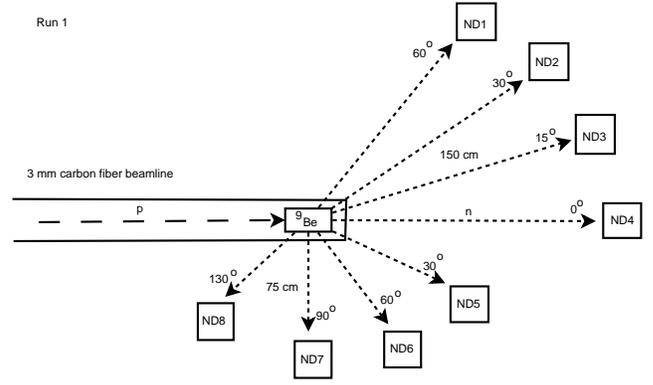}
\caption{\label{fig:setup}Schematic drawing of experimental setup in Run 1.}
\end{center}
\end{figure}

The neutrons produced in the target were measured by ToF technique. To this end RF signal from cyclotron was used as the reference time $t_{RF}$. Eight neutron detectors were installed simultaneously on the same height of the beamline at different angles and different distances $L$ around the target as shown for example in Fig.~\ref{fig:setup}.
Hence, the neutron kinetic energy was obtained from the hit time in the liquid scintillator $t_{hit}^n$
as following:
\begin{equation}
T_n=\frac{M_n}{2}\Biggl(\frac{L}{c(t_{hit}^n-t_{RF})}\Biggr)^2 ~,
\end{equation}
\noindent where $c$ is the speed of light and $M_n$ is the neutron mass.
The absolute value of the time difference $t_{hit}^n-t_{RF}$ was determined from the prompt $\gamma$ ToF.

\section{Data Analysis}
\subsection{DAQ Dead Time}
Each DAQ system stored the data from four independent detectors installed at different polar angles.
No on-line event pattern recognition or zero suppression were used in this experiment.
The identification of trigger channel was performed during off-line analysis.
The first in time detector hit, corresponding to largest TDC value above pedestal,
identified the trigger channel. Only the trigger channel data were selected in further
analysis. The other, non-trigger channels, were considered only for estimates of
multiple trigger events and for monitoring of pedestals. The probability of simultaneous hits in more
than one detector at a few kHz overall rate within 250 ns time interval covered by FERA Gate was negligible.
The cross talk between neighboring discriminator channels was generally below $10^{-3}$.
However, in some particular runs taken at small angles (0 or 5 degrees),
where the ratio of event rates in the two neighborhood detectors was very large and the average signal
amplitude in trigger channel high,
the cross talk reached 4\% for small detectors and 20\% for large detectors.
To correct for this effect the number of events having a secondary, non-trigger, TDC value
above TDC pedestal peak $N_{talk}$ was extracted for each run
by identifying the trigger hit as the one having the largest deposited energy.
Finally the live-time correction $L_{DAQ}$ was evaluated in each run as the ratio between
the number of recorded events $N_{event}$ and the number of trigger events recorded by the scalers $N_{trig}$
corrected for the cross talk events:
\begin{equation}
L_{DAQ}=\frac{N_{event}}{N_{trig}-N_{talk}}~,
\end{equation}
\noindent and the corresponding DAQ dead time:
\begin{equation}\label{eq:dead_time}
\tau_{DAQ}=\Bigl(1-L_{DAQ}\Bigr) \frac{t_{run}}{N_{trig}-N_{talk}}~.
\end{equation}
\noindent The obtained DAQ dead times varied from 11 to 16 $\mu$s,
a value compatible with FERA integration time.

From the measured total beam charge deposited on the target $Q_{beam}=\int_{t_{run}} I_{beam} (t) dt$
the total number of protons on target was obtained. This was performed correcting the collected charge
for DAQ live-time fraction $L_{DAQ}$:
\begin{equation}
N_p=\frac{Q_{beam} L_{DAQ}}{e}~.
\end{equation}
\noindent where $e$ is the elementary charge.

\begin{figure}[!t]
\begin{center}
\includegraphics[bb=2cm 6cm 20cm 23cm, scale=0.35]{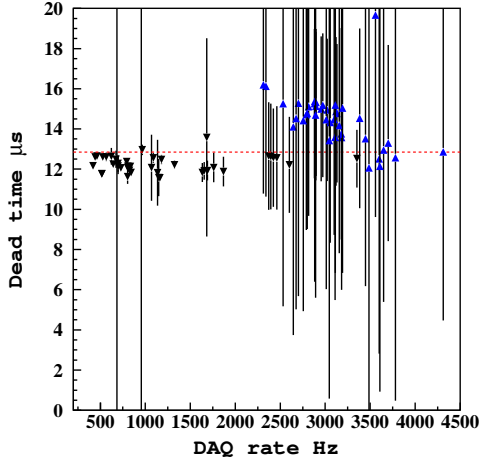}
\caption{\label{fig:daq_dead_time} DAQ dead time measured through Eq.~\ref{eq:dead_time}
as a function of DAQ event rate for large (black) and small (blue) detectors.
The dashed red line shows expected value 12.85 $\mu$s based on FERA manual.}
\end{center}
\end{figure}

\subsection{Neutron Identification}
\begin{figure}[!t]
\begin{center}
\includegraphics[bb=2cm 6.5cm 20cm 23cm, scale=0.35]{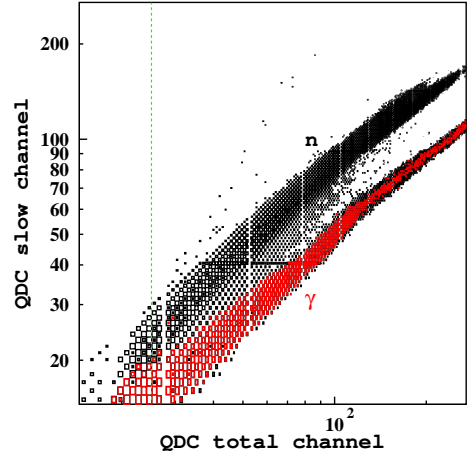}
\caption{\label{fig:psd_cuts_nd4} PSD distribution in the large detector obtained
with beam off (red), where only the $\gamma$ contribution is observed
and with beam on (black), where both species are present.
The green dashed line shows the minimal PSD threshold.}
\end{center}
\end{figure}

Neutron hits in the detectors were identified by means of PSD.
The determination of the optimal cut was performed by fitting the PSD distributions
for $\gamma$s and neutrons shown, for example, in Fig.~\ref{fig:psd_cuts_nd4}.
In fact, the data taken with beam off represent a pure sample of $\gamma$s,
allowing to determine the mean and the width of $\gamma$s PSD distribution.
To obtain the neutron PSD distribution we subtracted $\gamma$s distribution
from beam-on data.
Then, for each deposited energy we calculated the pulse tail value (QDC Slow)
lying in between $\gamma$ and neutron distributions which minimized the contamination
and inefficiency. Here we assumed both distributions being Gaussians.
The obtained set of points determined the optimal PSD cut curve.

\begin{figure}[!t]
\begin{center}
\includegraphics[bb=2cm 6.5cm 20cm 23cm, scale=0.35]{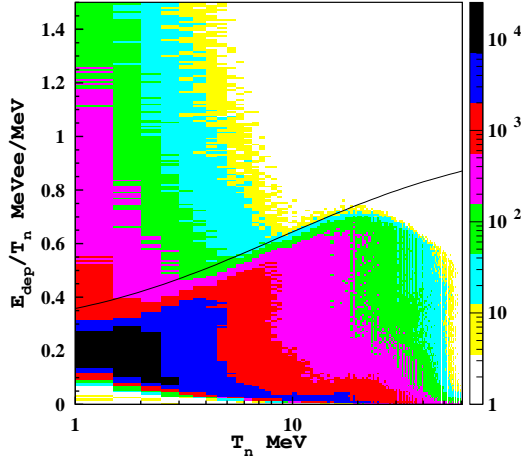}
\caption{\label{fig:edep_cuts} Energy deposited in the liquid scintillator as a function
of neutron kinetic energy measured in the large detector at 30 degrees.
The black line shows the parametrized maximum deposited energy $E_{dep}^{max}(T_n)$.}
\end{center}
\end{figure}

At small neutron energies the relative contribution of $\gamma$s in the total yield increases.
Moreover, the PSD becomes less efficient in rejecting them. An additional cut
on the correlation between the neutron energy reconstructed from Time-of-Flight
and deposited energy improves $\gamma$-rejection.
Indeed, the energy deposited in the liquid scintillator has to be smaller
than the maximum deposited energy corresponding to neutrons of given kinetic energy $E_{dep}^{max}(T_n)$.
The maximum deposited energy as a function of neutron kinetic energy was parametrized
from the data as shown in Fig.~\ref{fig:edep_cuts}.
Events with deposited energy above this maximum were rejected.

\subsection{Efficiency Evaluation}\label{sec:effic}
The efficiencies of neutron detectors were obtained by means of Geant4.9.5-p01
Monte Carlo simulations and verified in a dedicated measurement. 
Before making simulations we verified that the elastic and inelastic
cross sections of neutron scattering on hydrogen and carbon, implemented
in Geant4, were in agreement with ENDF.VII and EXFOR data.
We noticed that the use of G4CHIPSElastic model for $T_n>20$ MeV
requires the initialization of G4NeutronElasticXS database
to be called before HPElasticData. In the opposite case,
the lower precision G4NeutronElasticXS data substitute HPElasticData
in the entire kinematic range.

Measurements of the absolute efficiency were performed in the same experimental setup
used for detector calibrations. The technique developed in Refs.~\cite{Lanzano,Colonna} was employed.
This consists in a coincidence measurement of the neutron and $\gamma$ from the same fission.
Liquid scintillator neutron detectors were placed at the distance of 50 cm (100 cm for large cells)
from a spontaneous fission $^{252}$Cf source, while $\gamma$s were measured by a BaF$_2$ crystal, 
to which the fission source was directly attached.
Per each fission
$^{252}$Cf source emits $\nu=3.77$ neutrons and a large number of $\gamma$s ($>7$ with $E_\gamma>200$ keV~\cite{Verbinski:cf252g}). The BaF$_2$ crystal has a high efficiency for $\gamma$s, while its neutron efficiency
is an order of magnitude lower~\cite{Lanzano}. Moreover, applying PSD on the liquid scintillator signal
the probability to detect a $\gamma$ instead of the neutron drops at least by an order
of magnitude. Assuming that these contaminations are negligible, the absolute efficiency
of the liquid scintillator neutron detector is given by:
\begin{equation}
\epsilon_{ND}(T_n)=\frac{N_{coin.}(T_n)}{N_{BaF} A_{ND} \nu S_n(T_n)}~,
\end{equation}
\noindent where $N_{coin.}(T_n)$ is the number of measured coincidences
as a function of $T_n$, obtained from the time difference between liquid scintillator and BaF$_2$ hits,
$N_{BaF}$ is the total number of BaF$_2$ triggers, $A_{ND}$ is the
acceptance of liquid scintillator and $S_n(T_n)$ is the $^{252}$Cf neutron spectrum~\cite{cub}.
However, we noticed that the measured efficiency was increasing with BaF$_2$ threshold.
Because of relatively small contribution of the fast component in BaF$_2$ scintillator signal
for low deposited energy hits, the discriminator was perhaps triggering on the slow component,
almost 1 $\mu$s later.
Therefore we increased BaF$_2$ threshold value until saturation in the measured efficiency was observed.
The BaF$_2$ threshold energy at which saturation manifested itself was $E_\gamma \simeq 1$ MeV.
At such threshold the mean number of $\gamma$s per fission is about 2.

\begin{figure}[!t]
\begin{center}
\includegraphics[bb=2cm 6.5cm 20cm 23cm, scale=0.35]{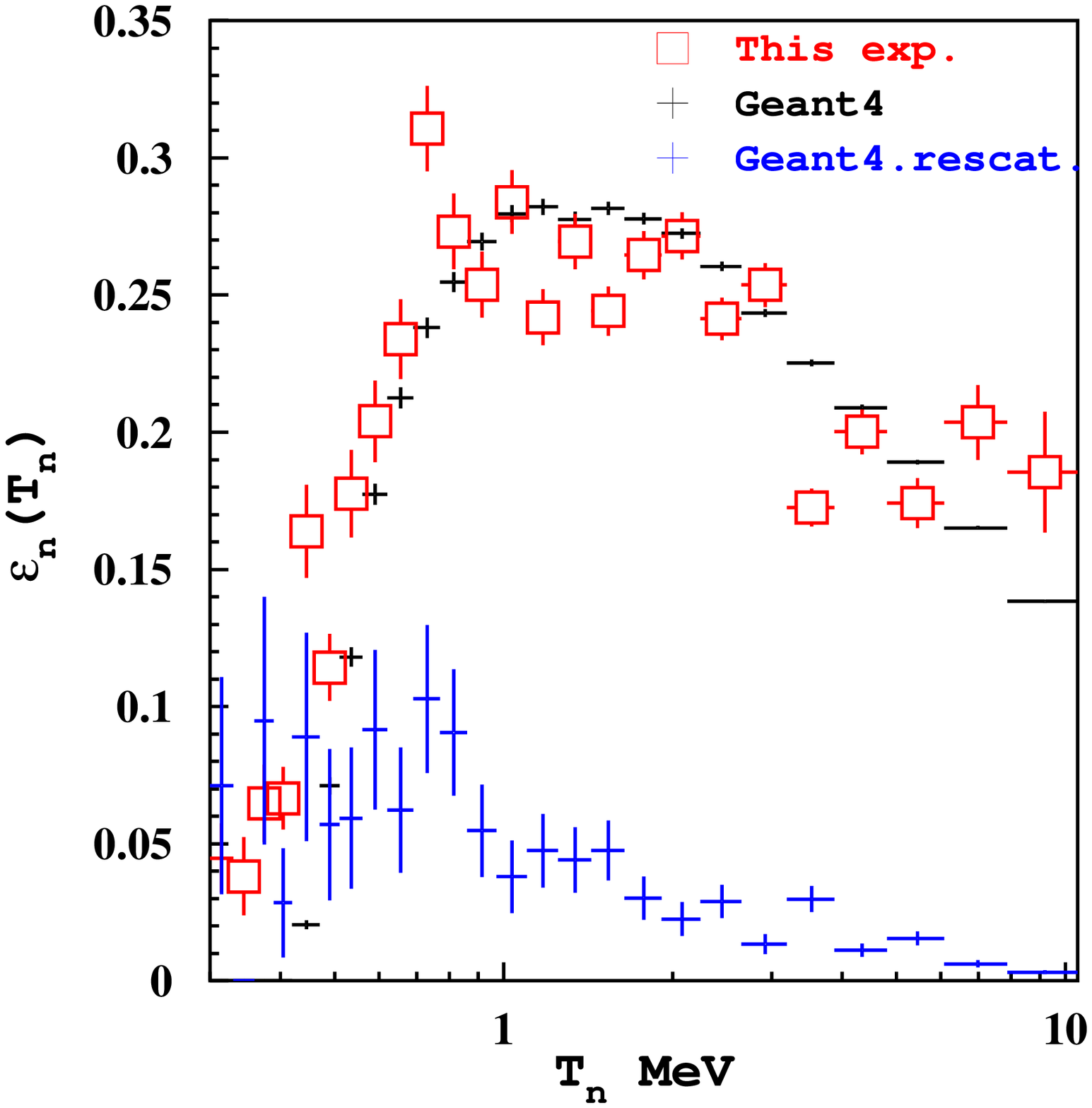}
\includegraphics[bb=2cm 6.5cm 20cm 24cm, scale=0.35]{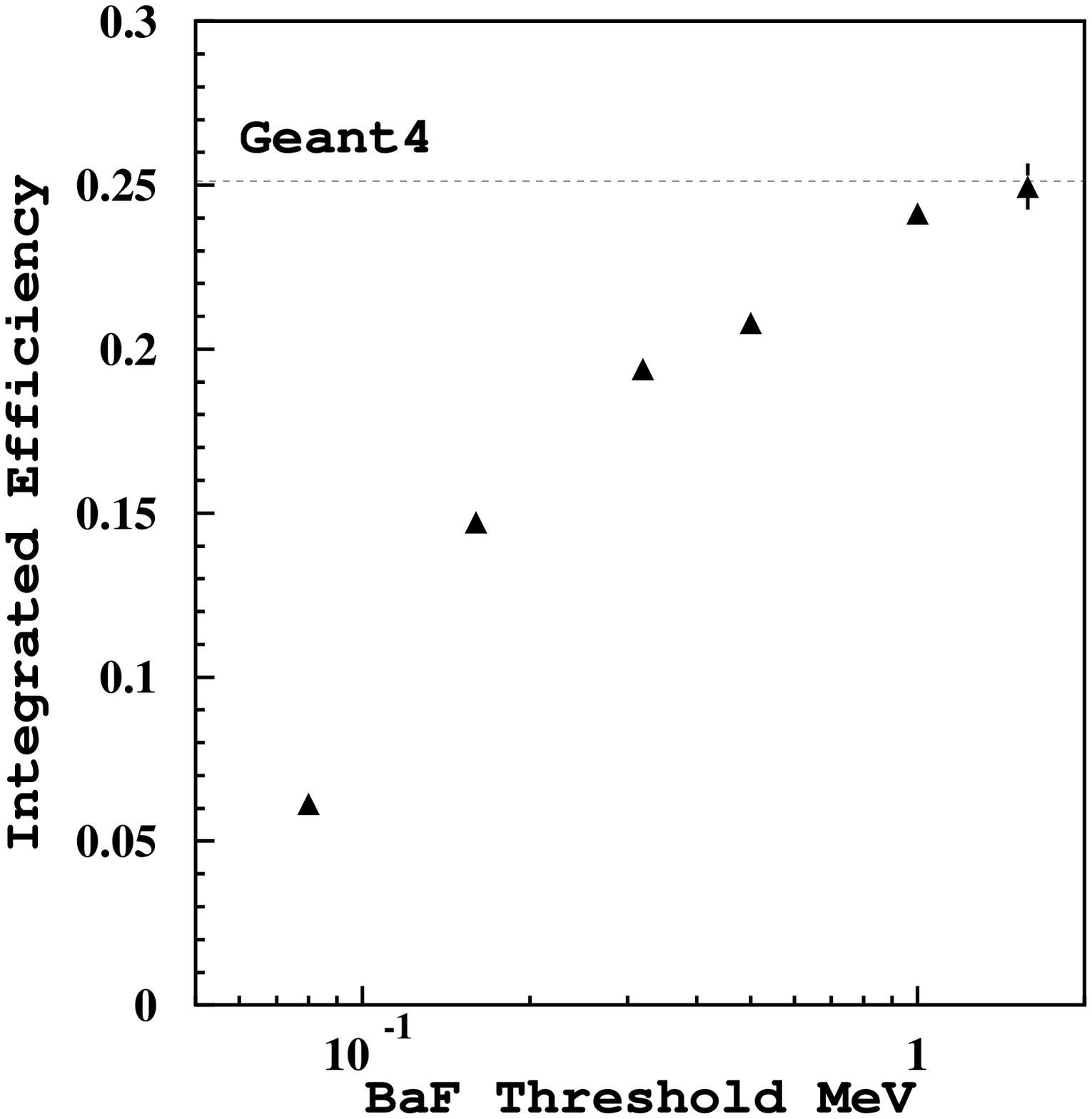}
\caption{\label{fig:g4_effs_small} Measured and simulated efficiency of large
liquid scintillator detectors (upper) and saturation of the integrated
efficiency as a function of coincidence detector (BaF) threshold.}
\end{center}
\end{figure}

The obtained efficiency for the small detector is shown in Fig.~\ref{fig:g4_effs_small}.
The measured efficiency was found to be in good agreement with Geant4 simulations over entire accessible
energy range, except for $T_n<0.7$ MeV.
We verified with further, more detailed, Geant4 simulations that the reason for this deviation was an incomplete
subtraction of the neutron rescattering from detector supports (shown in upper plot of Fig.~\ref{fig:g4_effs_small}
by the blue histogram).
The rescattering background was measured using a 50 cm long and 4.6 cm in diameter iron
shadow bar installed between $^{252}$Cf source and neutron detector.
However, the BaF$_2$ detector, used to detect $\gamma$s from $^{252}$Cf fission,
was installed on a fairly wide wooden table, rescattering from which
produced neutrons absorbed by the shadow bar. In this way the background
subtraction by means of the shadow bar was incomplete. The relative
contribution of this background drops rapidly with the neutron energy
and already at $T_n>1$ MeV it becomes smaller than systematic uncertainty.

In order to avoid large systematic uncertainties only energy region
where the efficiency exceeded 3\% was selected for the data analysis.

\subsection{Neutron Yield}
The two-fold differential neutron yield was extracted from the number of measured neutron hits
in the detector $N_n(T_n,\theta_n)$ as following:
\begin{equation}
\frac{1}{N_p}\frac{d^2N_n}{dT_n d\Omega_n}=
\frac{1}{N_p}\frac{N_n(T_n,\theta_n)}{\epsilon_n(T_n) \Delta T_n \Delta \Omega_n}~,
\end{equation}
\noindent where $N_p$ is the total number of incident protons on target,
$\epsilon_n(T_n)$ is the neutron detection efficiency obtained in Geant4 simulations,
$T_n$ and $\theta_n$ are neutron kinetic energy and emission angle with respect to the beamline, respectively.
$\Delta T_n$ is the neutron energy bin size and $\Delta \Omega_n$ is the detector solid angle.

In order to cover the entire measured energy range we combined the data from
small and large detectors. The maximum deposited energy measurable with small
detectors was 0.5 MeVee, corresponding to the maximum neutron energy of about 2 MeV.
Neutrons with energies above 2 MeV were also measured in small detectors,
but being their maximum deposited energy out of the DAQ range,
it was not possible to apply PSD and maximum deposited energy cut to the corresponding events.
From the other side the efficiency of large detectors drops rapidly below 2 MeV,
leading to a large systematic uncertainty on the observed yield.
Therefore, we selected small detector data at $T_n <2$ MeV and large
detector data at $T_n >2$ MeV.
In the overlap region two detectors demonstrate fairly good agreement
with average deviation of the order of 10\%.

\begin{figure}[!t]
\begin{center}
\includegraphics[bb=2cm 5.8cm 20cm 23cm, scale=0.35]{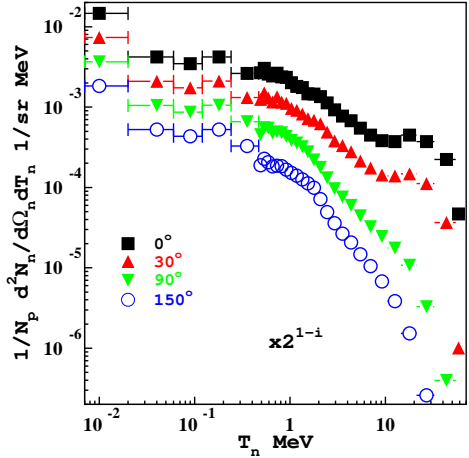}
\caption{\label{fig:eng_yield} Energy dependence of the neutron yield
for different production angles. Each successive angle distribution was
rescaled by a factor of 2 for visibility.}
\end{center}
\end{figure}

Examples of obtained yields are shown in Fig.~\ref{fig:eng_yield} as a function of neutron energy.
At low neutron energy the yields measured by liquid scintillators were extended
to the region $T_n<0.47$ MeV by means of $^3$He detector data described below.
The energy dependence is given by exponential fall-off with a bump-like
high energy shoulder observed at $\theta_n<60^\circ$.

\begin{figure}[!t]
\begin{center}
\includegraphics[bb=2cm 5.5cm 20cm 23cm, scale=0.33]{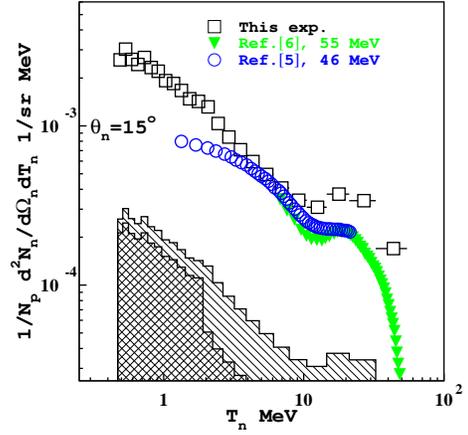}
\caption{\label{fig:cmp_world} Comparison of the measured neutron yield to world data from
Refs.~\cite{Johnsen76,Waterman79}.
The hatched histograms show systematic uncertainties due to
efficiency (right hatched) and normalization (left hatched).}
\end{center}
\end{figure}

The comparison of yields measured at 15$^\circ$ to world data at beam energies
nearby 62 MeV is shown in Fig.~\ref{fig:cmp_world}.
The data from Ref.~\cite{Johnsen76}, taken at somewhat lower beam energy,
though exhibiting similar structure are suppressed at high $T_n$ by the available phase space.
Instead, the data from Ref.~\cite{Waterman79} taken at even lower beam energy
have the low-energy behavior incompatible with our distribution.
%

\subsection{$^3$He and SiLiF Data}
The energy range $T_n<0.47$ MeV was measured by two different detectors:
50 cm long, 5 cm in diameter, 2 bar $^3$He tube (Canberra 150NH50/5A)
and 300 $\mu$m thick Micron MSX09 silicon detector attached to a plexiglass film
covered from detector side with 1.5 $\mu$m of $^6$LiF converter (SiLiF).
Signals from these two detectors were amplified by charge preamplifiers (ACHEM7F for $^3$He and Ortec 142A for SiLiF)
and then by Ortec 672 spectroscopy amplifier. The amplified pulses were
digitized by the same DAQ system using 12-bit SILENA 4418 peak sensing ADC.

These detectors were installed on the same supports used for liquid scintillators.
From liquid scintillator data we had seen that the low energy neutron flux is isotropic,
therefore we measured only few angles:
70 and 90 degrees for $^3$He tube and 130 degrees for SiLiF detector.
The distance to the target was chosen to be 64 cm for $^3$He tube
and 69 cm for SiLiF detector.

\begin{figure}[!t]
\begin{center}
\includegraphics[bb=2cm 6cm 20cm 23cm, scale=0.35]{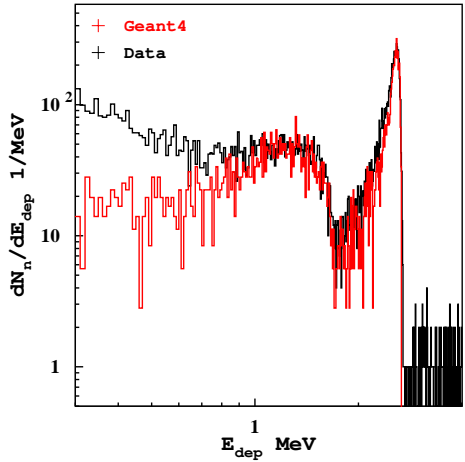}
\caption{\label{fig:lif_calib} SiLiF detector spectrum measured with thermalized AmBe source.
The black histogram shows the data, red histogram represents the Geant4 simulations.
The two peaks seen in these spectra are due to 2 MeV $\alpha$ and 2.7 MeV $t$ from the neutron conversion on $^6$Li.}
\end{center}
\end{figure}

The response of both detectors to thermalized neutrons from AmBe source was measured for the purpose of ADC calibration. The comparison of these calibration data to Geant4 simulations for SiLiF detector is shown
in Fig.~\ref{fig:lif_calib}. Except for the $E_{dep}<0.7$ MeV region, data and simulations were found
to be in good agreement. The enhancement in the data at $E_{dep}<0.7$ MeV is due to $\gamma$ background,
not included in the simulations.

\begin{figure}[!t]
\begin{center}
\includegraphics[bb=2cm 6cm 20cm 23cm, scale=0.35]{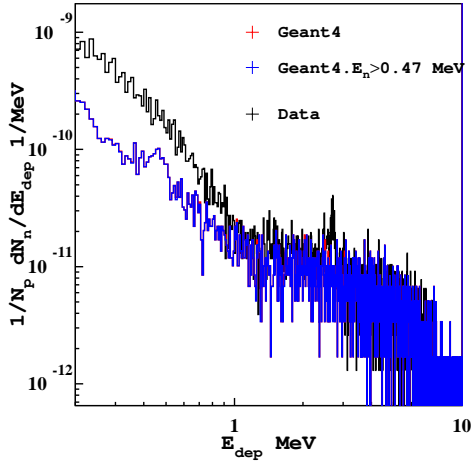}
\caption{\label{fig:lif_prod} Subtraction of the fast neutron contribution,
as measured by liquid scintillators ($T_n>0.47$ MeV), from the measured SiLiF spectra
using Geant4 Monte Carlo simulations.
The black histogram shows the SiLiF data, red histogram represents the Geant4 simulations of entire spectrum
and blue histogram gives the contribution of fast neutrons measured by liquid scintillators.}
\end{center}
\end{figure}

The SiLiF detector data taken during the experiment are shown in Fig.~\ref{fig:lif_prod}
in comparison to Geant4 simulations performed using the neutron yield measured by liquid scintillators.
In the energy range $E_{dep}>1$ MeV we obtain good agreement between data and simulations,
except for the small peak at 2.7 MeV in the data. This peak is likely to be due to the environmental
background of thermal neutrons, not measured for SiLiF detector.
In the low energy region we observe $\gamma$ background, similar to the one seen in calibration data.
Geant4 simulations showed that for the fast neutron spectrum of this experiment
the contribution of the neutron conversion in 1.5 $\mu$m $^6$LiF layer is negligible
with respect to the scattering off 300 $\mu$m silicon bulk.

$^3$He detector had unfavorable ratio of lateral surface to the detection window surface: about 40.
In order to reduce the environmental background it was wrapped in 1 mm thick Cd sheet.
The remaining background was measured by means of 50 cm long iron shadow bar with 1 mm thick
Cd sheet added at the end to absorb neutrons thermalized inside the shadow bar.
The obtained spectrum was found to be in good agreement with Geant4 simulations and with
liquid scintillator data. The yield obtained from $^3$He data was used in Fig.~\ref{fig:eng_yield}
to extend the covered energy range below 0.47 MeV.

\section{Conclusions}\label{sec:conclusions}
Two-fold differential neutron yield produced by 62 MeV proton beam off a thick, fully absorbing $^9$Be target was measured in complete energy range and almost complete angular interval.
Data precision is limited to about 10\% by systematic uncertainties on the detector efficiency and absolute normalization.

These data extend significantly the kinematic coverage of the world database in the 50-100 MeV beam energy range.
The data were found to be in good agreement with existing world data, except for Ref.~\cite{Waterman79}
which exhibited different low energy trend. More details on this experiment and
comparison of our data to MCNP, FLUKA and Geant4 simulations are given in Ref.~\cite{nima2013}.

The presented data will be used in the design of the ADS core~\cite{infn_e_ads} to describe
its neutron source.
These data can also be useful for the development of the neutrino source proposed recently in Ref.~\cite{isodar}.

\section*{Acknowledgment}
Authors would like to acknowledge excellent support provided during the experiment
by the accelerator staff and technical services of Laboratori Nazionali del Sud.
We also want to acknowledge the useful discussions with P.~Mastinu from Legnaro National
Laboratory, about the use of carbon fiber for the last beamline section.
This work was supported by the Istituto Nazionale di Fisica Nucleare INFN-E project.



\bibliographystyle{IEEEtran}
\bibliography{IEEEabrv,osipenko2013}
%

\end{document}